# Stacking of nanocrystalline graphene for Nano-Electro-Mechanical (NEM) actuator applications

**Kulothungan Jothiramalingam[1*], Marek E. Schmidt[1], Muruganathan Manoharan[1], Ahmed M.M. Hammam [1,2] and Hiroshi Mizuta[1,3]**

1   School of Materials Science, Japan Advanced Institute of Science and Technology, Nomi, Ishikawa 923-    1292, Japan.
2   Physics Department, Faculty of Science, Minia University, Main Road- Shalaby Land, Minia, 11432, Egypt
3   Hitachi Cambridge Laboratory, Hitachi Europe Ltd., J.J Thomson Avenue, CB3 0HE Cambridge, United
    Kingdom

[*]Correspondence: jothi@jaist.ac.jp;

**Abstract:** Graphene nano-electro-mechanical switches are promising components due to their excellent switching performance such as low pull-in voltage and low contact resistance. Mass fabrication with an appropriate counter electrode remains challenging. In this work, we report the stacking of nanocrystalline graphene (NCG) with a 70-nm dielectric separation layer. The buried NCG layer is contacted through the formation of vias and acts as actuation electrode. After metallization, the top 7.5-nm thin NCG layer is patterned to form double-clamped beams, and the structure is released by hydrofluoric acid etching. By applying a voltage between the top and buried NCG layer, a step-like current increase is observed below 1.5 V, caused by the contact of the movable beam with the buried NCG. No pull-out is observed due to the thin sacrificial layer and high beam length, resulting in low mechanical restoring force. We discuss the possible applications of the NCG stacking approach to realize Nano-Electro-Mechanical (NEM) contact switches and advanced logical components such as a AND logic.

## 1. Introduction

Nano-electro-mechanical (NEM) contact switches have received significant attention as a promising alternative to current complementary metal oxide semiconductor technology (CMOS) which increasingly suffer from high off-state leakage power due to the aggressive down-scaling (Lee et al. 2004; Dadgour and Banerjee 2007; Feng et al. 2010; Loh and Espinosa 2012; Peschot et al. 2015). The superior characteristics of the NEM contact switch such as near-zero off-state leakage current, high on-off ratios, CMOS technological compatibility, and steep switching slope (SS) less than the thermal limit of 60 mV/dec have been demonstrated (Lee et al. 2013). Materials for such NEM contact switches have to combine superior mechanical with good electrical properties. However, the available selection of materials is low (Loh and Espinosa 2012). Instead, integration of a dedicated contact material with silicon as a mechanical element, as it has been demonstrated for larger micro-electro-mechanical (MEM) switches, is often employed (Grogg et al. 2013). This approach is significantly more challenging for NEM switches as the geometric tolerances are more stringent. An encouraging alternative material is graphene, the atomically thin layer of carbon atoms with high Young's modulus of ~1TPa and high electrical conductivity (Geim and Novoselov 2007). Graphene is an ideal material for alternative switching devices such as tunneling field effect transistors (Hammam et al. 2017; Hamam et al. 2018). Graphene based NEM (GNEM) contact switches show low off-state leakage current, steep switching slope, and high on/off ratio (Milaninia et al. 2009; Sun et al. 2014, 2016a; Kulothungan et al. 2016). In spite of these inherent advantages, the GNEM contact switches are reported with high pull-in voltage (>1.5 V) and low switching reliability (Wagner and Vella 2013; Yaung et al. 2014; Sun et al. 2014; Jothiramalingam et al. 2017; Wang et al. 2017). In a typical GNEM contact switch, the fixed electrode is composed of metal. However, the irreversible surface adhesion (stiction) can occur between graphene and the metal electrode due to the molecular covalent bond formation (Cheng et al. 2011). Thus, a different contact pair is required. Nanocrystalline graphene (NCG) that can be grown at wafer-scale with variable thickness using plasma-enhanced chemical vapor deposition (PECVD) has been demonstrated to be a viable material for two-terminal NEMS switches with metal electrodes (Sun et al. 2016b; Fishlock et al. 2016). To enable the use of the metal-free growth of NCG for switches without metal, a newly engineered approach is required, that is compatible with the relatively high deposition temperature.

In this work, we report the stacking of NCG with a thin dielectric separation layer. We demonstrate contacting of the buried NCG layer through the formation of vias, and the pull-in of a movable double-clamped beam onto the buried layer. We discuss how the stacked NCG could be used to realize three-terminal switches based on triple-stacked NCG, and AND logic with a single movable beam.

**2. Experimental**

*2.1 Stacked NCG-based device fabrication*

Figure 1 illustrates the fabrication process of the stacked NCG-based device with the movable beam. First, a NCG film is deposited on Si/SiO$_2$ (285 nm) by PECVD method as shown in Figure 1a (deposition temperature, 850°C; duration, 15 min; RF power, 100 W; CH$_4$:H$_2$ ratio, 0.8:1). The thickness of this NCG film is ~45 nm. The high resolution negative resist HSQ (hydrogen silsesquioxane), is spin-coated with a thickness of ~70 nm on the NCG film as illustrated in the Figure 1b. After baking at 650°C for 30 min, the HSQ layer is fully converted to SiO$_2$ and acts as the sacrificial layer, which determines the airgap between the bottom NCG and the movable NCG film. Then, the second layer of NCG is grown on top of the SiO$_2$ layer with ~7.5 nm thickness (deposition temperature, 800°C; duration, 7 min 30 sec; RF power, 100 W; CH$_4$:H$_2$ ratio, 0.8:1; Figure 1c).

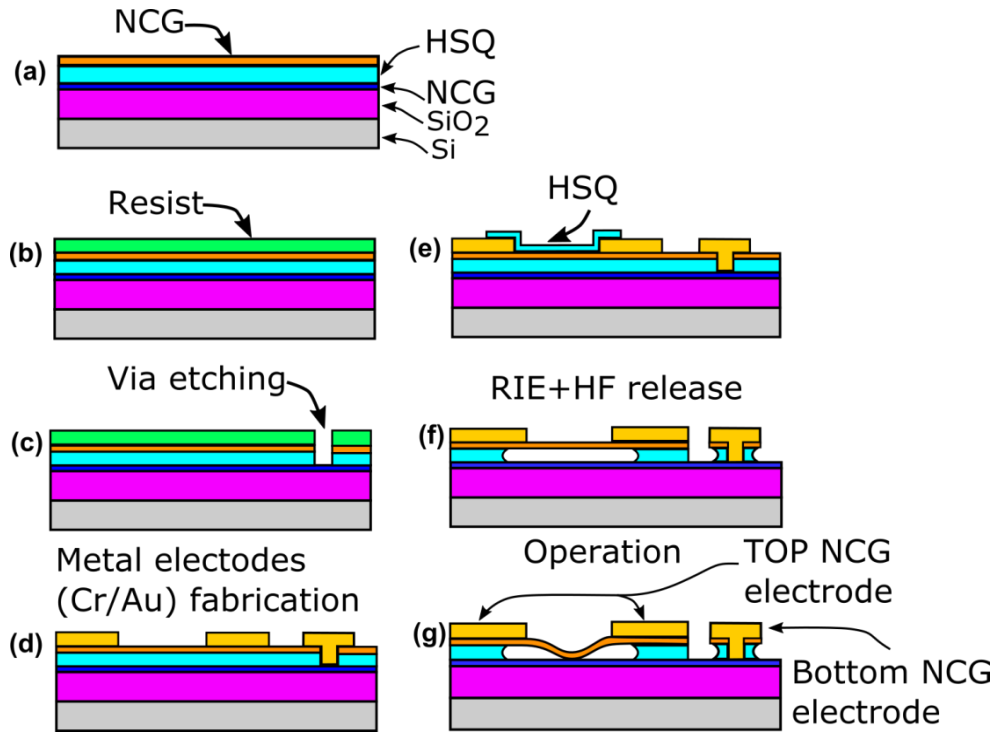

Figure 1. Schematic illustration of stacked-NCG based device fabrication with the movable beam.

Next, openings through the top NCG and the $SiO_2$ sacrificial layer are prepared to allow contacting of the bottom layer. Small squares, which can be as small as 500x500 nm$^2$, are defined using Poly (methyl methacrylate) (PMMA) resist and electron beam lithography (EBL, Figure 1b and 1c). First, the top NCG layer is etched by reactive ion etching (RIE; $O_2$ flow, 20 sccm; pressure, 2.6 Pa; RF power, 20 Watt; duration, 3 min), followed by RIE etching of the $SiO_2$ layer. The etch conditions are carefully adjusted to ensure a full opening of the vias while avoiding excessive etching of the buried NCG layer. Here, $O_2$ gas (2 sccm) and $CHF_3$ (40 sccm) are used at a moderate RF power of 50 Watt and 4 Pa chamber pressure. After 4 minutes of etching, the via is fully opened. Note that here the buried NCG is very thick, and slight over etching does not affect device operation. Contacts to the top and bottom NCG layers are realized by Cr/Au (5 nm/85 nm) metal electrodes. Chromium is chosen over titanium as adhesion layer since it withstands the HF release step. Methyl methacrylate (MMA)/PMMA bilayer resist is patterned by EBL, and the metal films are deposited using electron beam evaporation (Figure 1d). To pattern the top NCG, another layer of HSQ is spin coated and patterned by EBL (MF319 developer) (Figure 1e). Next, the exposed top NCG is etched with oxygen RIE (identical conditions). Finally, the sacrificial $SiO_2$ layer and the HSQ etch mask is etched in buffered hydrofluoric acid (1:5) for 30 seconds to release the top NCG film and form the suspended NCG beam. The device is dried using a supercritical point dryer to prevent the capillary force induced collapse of the suspended graphene beams. The suspended structure of the stacked NCG-based device with movable beam is shown in Figure 1f, and Figure 1g illustrates the deflection of the beam onto the bottom NCG layer. Several devices were fabricated by this method, and the top NCG layer was patterned to comprise either a single double-clamped NCG beam or an array of double-clamped beams with different length (all connected to the identical electrode).

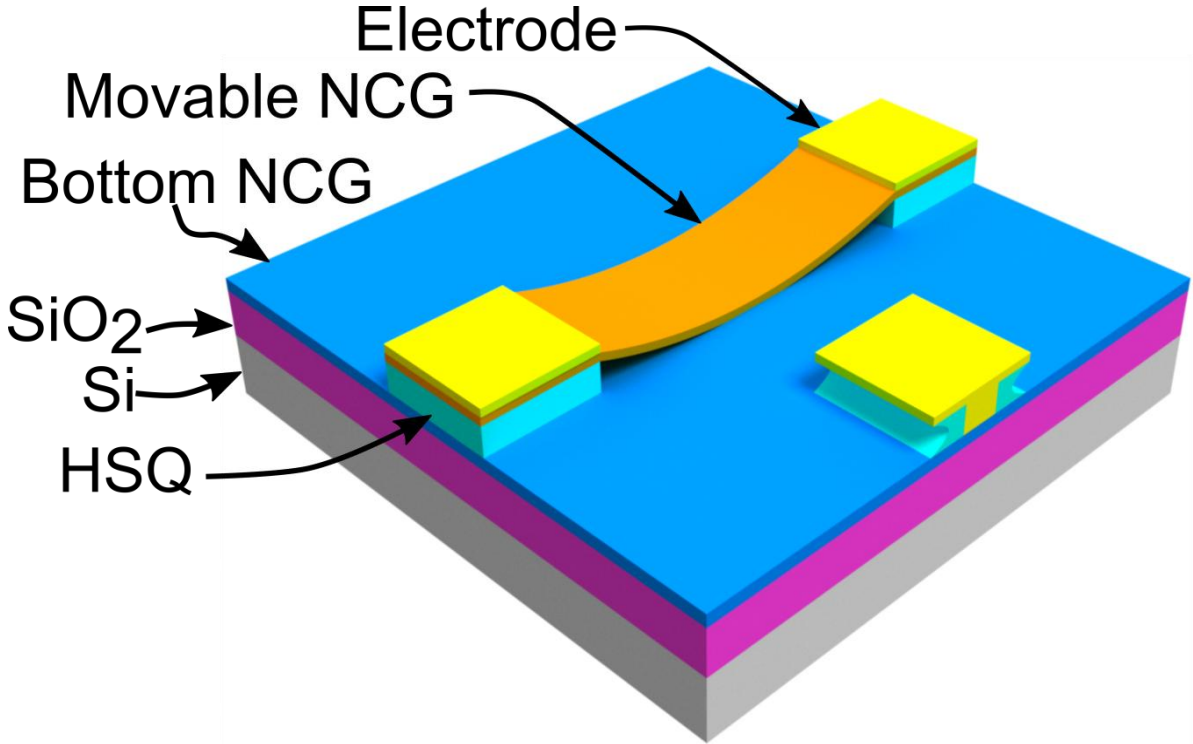

Figure 2. Schematic of two-terminal stacked NCG-based devices with the movable beam.

*2.2 Electrostatic actuation of stacked NCG-based devices*

A stacked NCG-based device with movable NCG beam is illustrated in Figure 2 in the pulled-in state. When a voltage is applied between the movable NCG and bottom NCG, an electrostatic force acts between the two conductive layers corresponding to the potential difference and distance. As the bottom NCG is fixed to the substrate, this force causes a mechanical deflection of the movable beam, which results in a mechanical restoring force that depends on the dimensions and properties of the beam. When the applied voltage reaches a critical point, the electrostatic force overwhelms the restoring force, which causes the NCG beam to be pulled towards the bottom NCG. The abrupt deformation of the movable NCG beam onto the bottom NCG causes the formation of an electrical contact, and a sharp rise in the current between the deformed NCG and bottom NCG is observed. The pull-in voltage can be estimated analytically for the case of a double-clamped NCG beam by

$$V_{pi} = \sqrt{\frac{8kg_0^3}{27\varepsilon_0 W_b W_c}} \qquad (1)$$

$$k = 32EW_b \left(\frac{t}{L}\right)^3 \qquad (2)$$

where $g_0$ is the initial air gap between the double-clamped NCG beam and the bottom NCG, $k$ is the beam spring constant, $\varepsilon_0$ is the vacuum permittivity of $8.853 \times 10^{12}$ F·m$^{-1}$, $L$, $t$, and $W_b$ are the NCG graphene beam length, thickness, and width, respectively. $W_c$ is the contact width, $E$ is Young's modulus of NCG.

Electrical characterization is performed in vacuum condition (~0.1 Pa). Measuring the devices in vacuum condition reduces the molecular absorption. First, the electrical conduction through the double-clamped NCG beam is measured in a reference device as shown in Figure 3a. The metallic response is visible in the current-voltage (I–V) measurements in Figure 3b (16.85 kΩ), signifying ohmic contact between the NCG and the Cr/Au electrodes. Next, the two-terminal measurement configuration is used to study the actuation of the movable beam (Figure 3c). The voltage is applied between the bottom NCG layer (through the via) and the NCG beam's contact electrode. The applied voltage is swept from zero volts to high voltages (Keithley 4200-SCS semiconductor device analyzer) while monitoring the current.

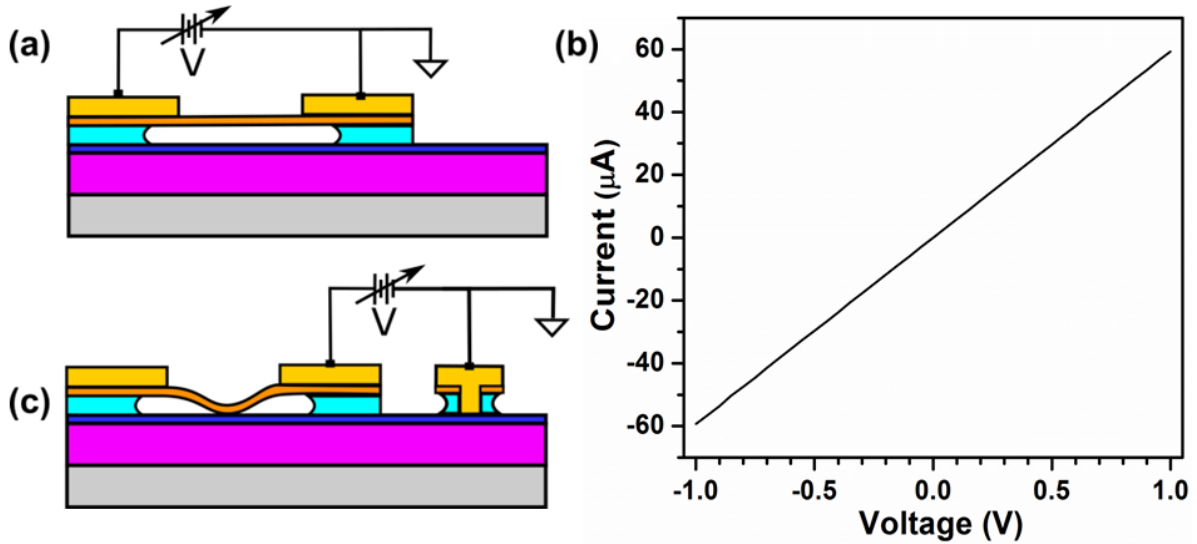

Figure 3. (a) Measurement configuration for the I-V response of the NCG beam. (b) I-V response of suspended NCG beam measured between the contact electrodes in a reference device. (c) Two terminal configurations to measure actuation of the stacked NCG-based device with the movable beam.

## 3. Results and Discussion

*3.1 Electrical characteristics of the stacked NCG-based devices*

Firstly, we discuss the actuation characteristics of single beam NCG-NCG devices. A scanning electron microscope images of a fabricated stacked NCG-based device with single double-clamped NCG beam is shown in Figure 4a after pulled-in. The bottom NCG film is used as the contact material, and the NCG beam has a length, $L$, of 1.35 µm and a width, $W$, of 0.25 µm. The initial airgap between the movable NCG beam and the buried NCG was ~70 nm, as defined by the SiO$_2$ sacrificial layer thickness.

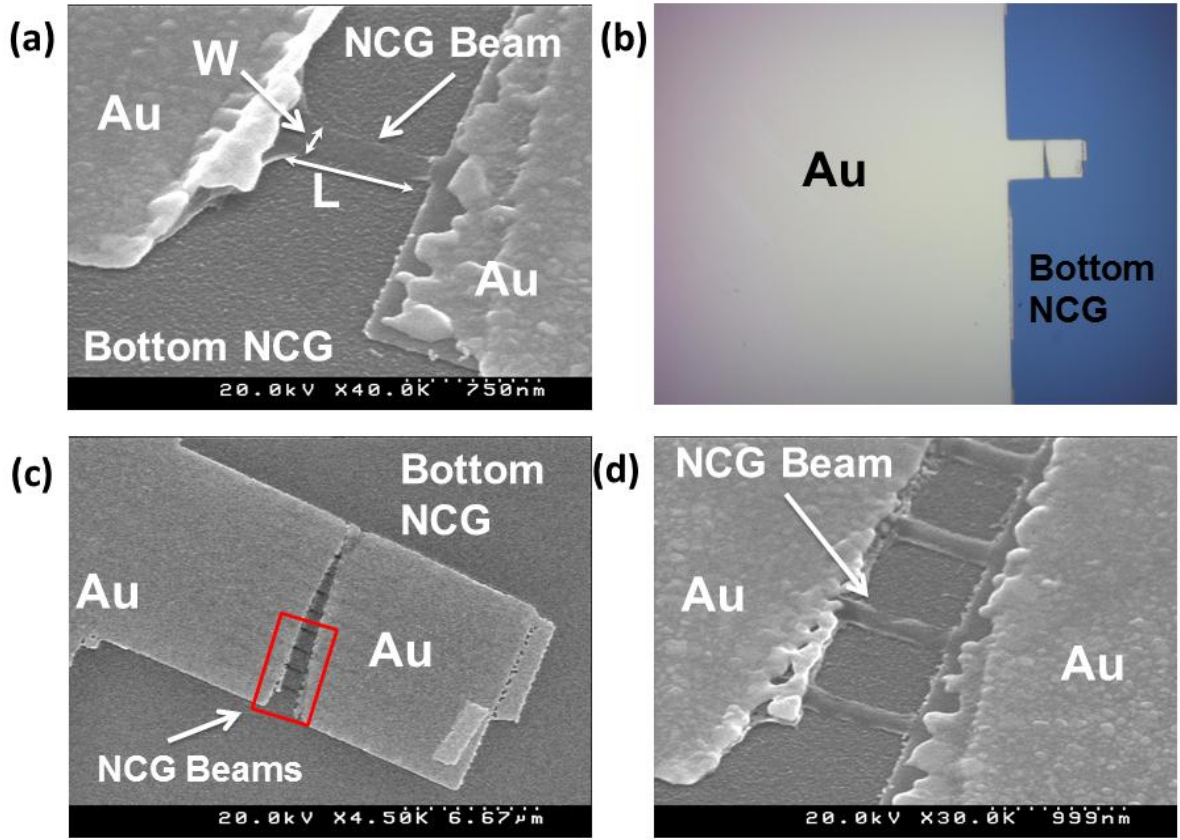

Figure 4. (a) Optical microscope image of the stacked NCG-based device with movable NCG beam. (b) A single stacked NCG-based device after pull-in. (c) An array of double-clamped movable NCG beams. (d) Zoom-in view of the NCG ribbons in (c).

The electrical characterization of the two-terminal actuation of the single beam NCG-NCG device from Figure 4a is shown in Figures 5a. The voltage on the x-axis represents the applied voltage between the movable NCG beam and the bottom NCG film (compare Figure 3c). The values on the y-axis denote the current between the contact electrode and the via electrode. The actuation of the stacked NCG-based device with only a single movable beam is shown in Figure 5a. At around 1.4 V, a remarkably sharp increase in current is observed, saturating at the compliance value of 50 µA. The current compliance was set to 50 µA to avoid the device failure due to the Joule heating. Unfortunately, no pull-out is observed, and the second scan of voltage shows ohmic contact between the movable beam and the bottom NCG with a resistance of ~13.6 kΩ. From Figure 4b, it is visible that the movable NCG beam is collapsed onto the bottom NCG. The reason for the lack of pull-out is ascribed to the following factors: (i) the movable NCG is only 7.5 nm thick, therefore it has a low mechanical stiffness; (ii) the air gap of only 70 nm together with the length of 1.35 µm does not create enough mechanical restoring force; (iii) weak compressive strain in the NCG further increases the required gap to generate sufficient restoring force.

The mechanical restoring force of a double-clamped beam with spring constant k (Equation (2)) is given by

$$F_{restoring} = -kd \quad (3)$$

where *d* is the deflection. Therefore, the restoring force could be significantly increased by using thicker dielectric during the NCG stacking process or reducing the beam length, similar to previous reports of NCG-based NEM switches *(Sun et al. 2016b)*.

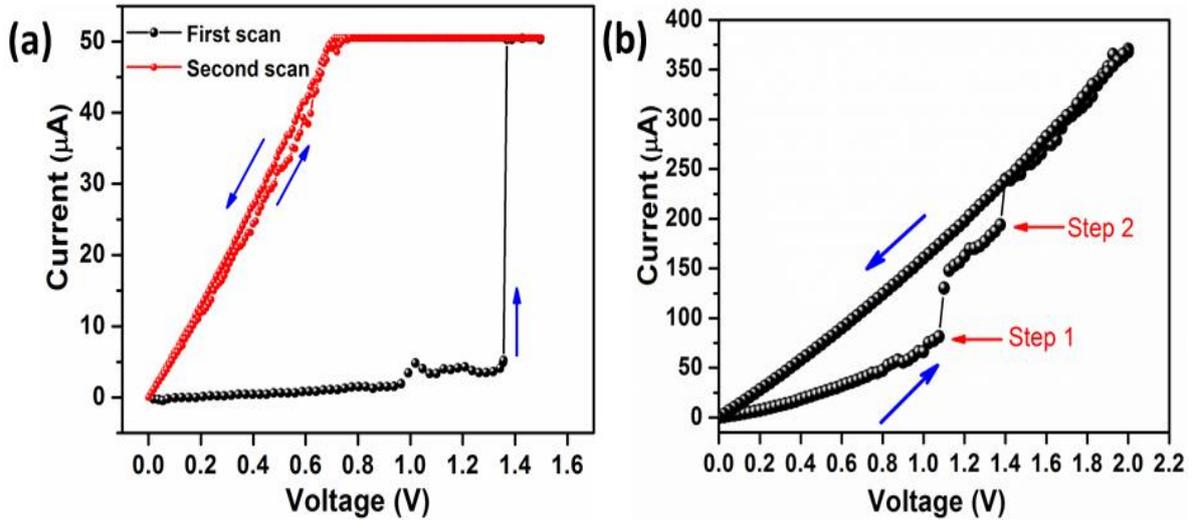

Figure 5. Actuation characteristics of the stacked NCG-based device with movable NCG beam. (a) Device with a single beam (*L* = 1.35 μm, *W* = 0.25 μm) showing clear pull-in at ~1.35 V. (b) Device with multiple beams. Initially, some of the beams are in contact with the bottom NCG. At ~1.03 and ~1.35 V, current jumps are observed, signifying pull-in of additional beams.

Next, we discuss the operation of stacked NCG-based devices with an array of movable double-clamped NCG beams. Figure 4b shows an optical microscope image of a typical array type device, where the array of NCG beams are located between the large pad and the small anchor island. Figure 4c and d illustrates the array of NCG double-clamped beams with a width, *W*, of 0.25 μm and gradually decreasing length from 1.35 μm to 0.4 μm. The array type stacked NCG-based device comprises eight double-clamped beams. The actuation characteristics of the array type device is shown in Figure 5b. The array type device exhibits a relatively large ohmic-like leakage current. The slight non-linearity is due to the weak contact between the movable and the bottom electrode; the full potential difference is not present between the elements due to the leakage current, however, due to the non-negligible contact resistance, the contact becomes increasingly better with higher voltage. At a two-terminal voltage of ~1.03 V and ~1.35 V, sharp increases of current are observed. When reducing the voltage back to 0 V, no steps are visible. This signifies that, after suspension, some of the double-clamped beams were already in contact with the bottom NCG creating a leakage path. Two of the beams, which were suspended previously, were successively pulled-in by the applied voltage. The two different pull-in voltages, i.e., at ~1.03 V and ~1.35 V are from different NCG beams. The resistance contribution from these two beams can be calculated by assuming the circuit of parallel resistors. Before pull-in, one of the resistors is equal to the device resistance, and the other resistor (representing the contact resistance of the suspended beam) is assumed to be infinite. With the two resistance steps, $\Delta R$ of 6.5 k$\Omega$ (Step 1) and 1.5 k$\Omega$ (Step 2), the resistance contribution of these two additional conduction channels can be calculated. The values are 17.2 k$\Omega$ and 26.5 k$\Omega$, which is of similar magnitude with the device in Figure 5a. In the reported devices, the distance between the via and the contact area is ~250 µm. Therefore, not all of the applied voltage is actually present at the contact area as some of the potential is lost in the bottom NCG film. This effect can be reduced by patterning the vias within a few microns of the movable beam.

*3.2 Proposed advanced switching devices using stacked NCG-based devices*

NCG has very good adhesion to SiO$_2$. Therefore, NCG can be patterned by O$_2$ RIE into any shape without risking peeling during the remaining fabrication. Also, masked RIE etching can create finer pattern than metal deposition by lift-off. Furthermore, a third SiO$_2$/NCG stack could be added. This provides for some interesting possibilities in terms of device design which we want to discuss briefly, given that pull-out is realized by appropriate device geometry optimization.

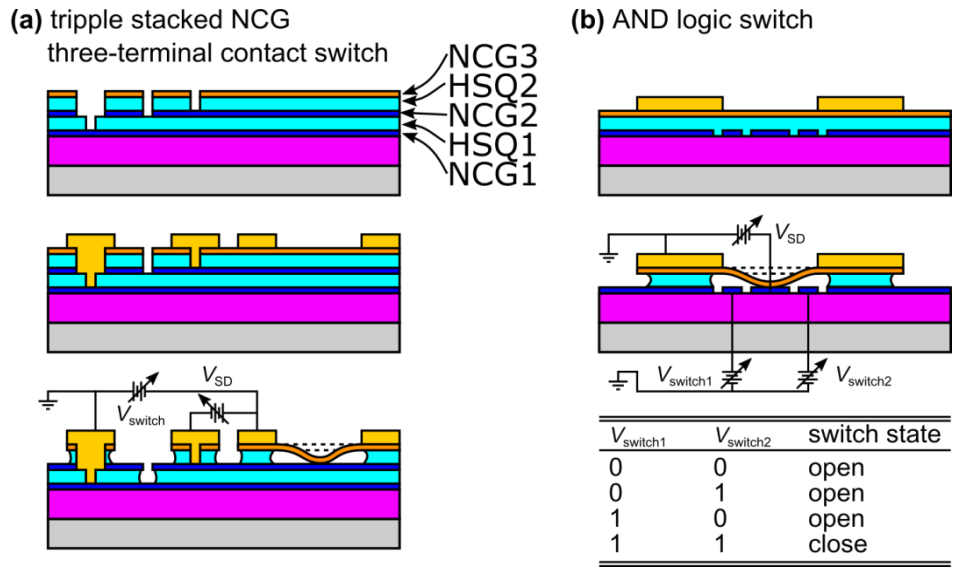

Figure 6. Proposed NCG based three terminal NEM contact switch. (b) AND logic using the NCG NEM switch.

Often, two-terminal devices are realized due to their simplicity. However, they are not always desired as the current through the switch is dependent on the switch resistance and pull-in voltage, although this effect can be addressed in the CMOS circuit design. Nevertheless, by using the conductive silicon substrate as the actuation electrode, switching can be performed while the voltage across the switch is set to zero. Once switching occurred, the voltage and current across the switch can be set as required, independent of the switching requirement. The drawback of using the silicon substrate for switching is that the switches on the same substrate can only be operated independently if the substrate is grounded. In such case, the interconnection of switches is not possible as their potentials will vary. Thus, in Figure 6a, we show an approach to realize three-terminal switches in triple-stacked NCG with individual switching electrodes. The bottom most NCG layer is pre-patterned before stacking (not shown), and vias are formed to contact the actuation gate, which is individual to a single switch, and the inner NCG layer. The switch itself in the second and third NCG layer is identical to the one shown in Figure 1 and 2. Due to the electrostatic screening of the second NCG layer, an appropriate planar geometry has to be ensured.

Another possibility, which is opened by pre-patterning the NCG before applying the next layer of $SiO_2$ and the NCG growth, is a AND logic switch with a single movable beam (Figure 6b). The movable beam is suspended above three individual NCG electrodes in the bottom NCG layer, of which the middle one is the contact electrode. The left and right electrodes, which can be addressed individually through vias, exert an electrostatic force on the movable beam. By careful geometric design of the switch, it should be possible to cause switching only if a high bias is applied to both switching electrodes. Of course, such switch can also be configured as a three-terminal switch by connecting the two switching electrodes.

## 4. Conclusions

We reported stacking of nanocrystalline graphene with a dielectric separation layer. This stack is used to realize devices with a fixed bottom NCG electrode and a movable NCG beam. A sharp current increase across the NCG films is observed when the potential exceeds 1.5 V due to the electrostatic pull-in of the movable NCG beam. A device with an array of movable beams is reported as well, and the pull-in characteristics are used to extract contact resistance values between 16 kΩ and 27 kΩ. Issues arising from the leakage current and the large distance of the via from the movable beams are discussed, as well as the requirements to achieve pull-out. Advanced device architectures, such as triple-stacked NCG-based three-terminal switches and a NEM based AND logic switch with a single movable beam are proposed. With the possibility of large area metal-free NCG growth, the stacking-approach is useful to exploit the material. In future, the device geometry and the NCG stacking will be optimized to achieve reliable pull-out while maintaining low pull-in voltage.

**Acknowledgments**: This work was supported by the Grant-in-Aid for Scientific Research No. 25220904, 16K13650, and 16K18090 from Japan Society for the Promotion of Science and the Center of Innovation (COI) program of the Japan Science and Technology Agency. The authors thank Wenzhen Wang for the experimental assistance, and Harold M. H. Chong, Zaharah Johari and Jamie Reynolds for the assistance with NCG growth.